\documentclass[11pt,a4paper]{article}
\pdfoutput=1
\usepackage{jheppub}
\usepackage{gensymb}
\usepackage{placeins}
\usepackage{subfigure}
\usepackage{amssymb,amsmath}
\usepackage{graphicx}
\usepackage{color}
\usepackage{cancel}
\usepackage[colorlinks=true
,urlcolor=blue
,citecolor=blue
,linkcolor=blue
,pagecolor=blue
,linktocpage=true
,pdfproducer=medialab
]{hyperref}

 \usepackage[numbers]{natbib}
\usepackage{notoccite}
\makeatletter \renewcommand{\@dotsep}{10000} \makeatother
\def\be{\begin{equation}}
\def\ee{\end{equation}}
\def\bea{\begin{eqnarray}}
\def\eea{\end{eqnarray}}
\def\bi{\begin{itemize}}
\def\ei{\end{itemize}}

\usepackage[nodisplayskipstretch]{setspace}

\newcommand{\PS}{SU(4)_{c}\times SU(2)_{L}\times SU(2)_{R}}
\newcommand{\mgut}{M_{{\rm GUT}}}

\DeclareUnicodeCharacter{2212}{-}
\DeclareUnicodeCharacter{202F}{\,}
\graphicspath{{/Users/cemsalihun/Desktop/PSLRg2/Lowg2/plot/tex}}

\begin{document}

\begin{titlepage}
\pagestyle{empty}

\vspace*{0.2in}
\begin{center}
{\Large \bf Muon $g-2$ and dark matter in Supersymmetric $SU(4)_c \times SU(2)_L \times SU(2)_R$} \\
\vspace{1cm}
{\bf Qaisar Shafi$^{a,}$\footnote{E-mail: qshafi@udel.edu}, Amit Tiwari$^{a,}$\footnote{E-mail: amitiit@udel.edu} and
Cem Salih $\ddot{\rm U}$n$^{b}$\footnote{E-mail: cemsalihun@uludag.edu.tr}}
\vspace{0.5cm}

{\small \it
$^a$Bartol Research Institute, Department of Physics and Astronomy,
University of Delaware, Newark, DE 19716, USA\\
$^b$Department of Physics, Bursa Uluda\~{g} University, TR16059 Bursa, Turkey}

\end{center}

\vspace{0.5cm}
\begin{abstract}

The latest FermiLab muon $g-2$ result shows a $5\sigma$ discrepancy with a ``widely advertised" Standard Model prediction. We consider a supersymmetric $SU(4)_c \times SU(2)_L \times  SU(2)_R$ model in which this discrepancy is resolved by including contributions to muon $g-2$ from a relatively light SUSY sector. A variety of realistic coannihilation scenarios can reproduce the observed dark matter relic abundance. With a significantly reduced discrepancy, of order $1 \sigma$ or less, the Higgsino-like dark matter solutions are also viable. We provide benchmark points for these solutions that will be probed in the direct detection dark matter experiments and collider searches.

\end{abstract}
\end{titlepage}


\section{Introduction}
\label{sec:intro}
The supersymmetric extension of the well-known gauge symmetry $SU(4)_c \times SU(2)_L \times SU(2)_R$ \cite{Pati:1974yy}, for short 4-2-2, has been extensively utilized to explore various phenomenological implications including t-b-tau Yukawa unification, sparticle spectroscopy, dark matter candidates and muon $g-2$ (see, for instance, \cite{Gogoladze:2009ug,Gogoladze:2009bn,Gogoladze:2010fu,Gogoladze:2012ii,Raza:2014upa,Shafi:2015lfa,Altin:2017sxx}). In a recent study \cite{Shafi:2023sqa}, we investigated third family Quasi-Yukawa unification (QYU) and identified coannihilation scenarios that yield the desired dark matter (DM) relic abundance. We also found solutions compatible with QYU that yield TeV scale Higgsino dark matter.

This paper is stimulated by the latest results on muon $g-2$ presented by the Fermilab collaboration (FNAL) \cite{Muong-2:2023cdq}, which has updated the World average to \( a_\mu(\text{Exp}) = 116 592 059(22) \times 10^{-11} \) (0.19 ppm). This measurement shows a \( 5.1\sigma \) discrepancy from the widely advertised Standard Model (SM) predictions \cite{Aoyama:2020ynm}, expressed as

\begin{equation}
\Delta a_{\mu} \equiv a_{\mu}^{{\rm exp}} - a_{\mu}^{{\rm SM}} = (24.5 - 4.9)\times 10^{-10}.
\label{eq:damu}
\end{equation}
This strongly indicates the possible presence of new physics beyond the SM \cite{Alok:2023bfk,Abdallah:2023pbl,Chakraborti:2023pis,Lu:2021vcp,Chun:2021rtk,Das:2021zea,Cadeddu:2021dqx,Criado:2021qpd,Chakrabarty:2020jro, Paul:2023qkw, Gomez:2023syh,SalihUn:2023unj}, and TeV scale supersymmetry remains an attractive leading candidate for this role. The supersymmetric 4-2-2 model we consider here for additional contributions to muon $g-2$ has a number of attractive features. It incorporates electric charge quantization, predicts the existence of heavy right handed neutrinos, and accommodates the observed tiny neutrino masses through the seesaw mechanism. If the lightest supersymmetric particle (LSP) is stable, electrically neutral and weakly interacting, it becomes a plausible DM candidate. We look for solutions in this 4-2-2 model that resolve the muon $g-2$ anomaly and are compatible with the bound on the DM relic abundance reported by the Planck satellite experiment \cite{Planck:2018nkj}. If the muon $g-2$ discrepancy is confirmed to be around the $5 \sigma$ level, the viable DM scenarios we obtain rely on coannihilations between the LSP and one or more NLSPs (next to LSP in mass.) 

The SM predictions, despite uncertainties associated with strong interactions \cite{Cvetic:2020unz,Chakraborty:2018iyb,Izubuchi:2018tdd}, stand in contrast with a lattice QCD result by the BMW collaboration, which is \( a_{\mu}(HVP; BMW) = 711(18)\times10^{-11} \) \cite{Borsanyi:2020mff}, aligning more closely with the experimental average. Further studies are underway to clarify these theoretical differences \cite{Colangelo:2022jxc}. If the muon $g-2$ discrepancy is significantly lower, less than or of order $1\sigma$ the higgsino DM solution also becomes viable and these solutions are also compatible with third family QYU.

In this paper, we assume the presence of a flavor symmetry along with the $\PS$ group at the grand unification scale ($\mgut \simeq 2\times 10^{16}$ GeV), which helps split the third family  from the first-two families (for a detailed discussion see \cite{Babu:2014sga}). In our calculations we assume that `C-parity' \cite{Kibble:1982dd,Lazarides:1985my}, which interchanges left and right, and conjugates the representation, is broken at $\mgut$. Somewhat loosely we call it left-right (LR) symmetry, and its breaking allows us to assume different soft supersymmetry breaking gaugino and scalar masses in the left and right sectors of the model. 

The prior analyses on muon \( g-2 \) within this class of SUSY GUTs have demonstrated that solutions compatible with the muon \( g-2 \) measurements within \( 1-2\sigma \) have significant implications. Notably, these constraints exclude Higgsino-like DM solutions~\cite{Shafi:2021jcg,Shafi:2023sqa}. This leads to a SM-like Higgs boson mass of approximately \( 123 \) GeV~\cite{Gomez:2022qrb}, which is marginally acceptable given the theoretical uncertainties. It also suggests relatively light sleptons with masses \( \lesssim 600 \) GeV, among other outcomes. However, the recent Fermilab results for muon \( g-2 \) imply that the previously established bounds can be expanded based on the new average, and scenarios previously ruled out by the \( 1-2\sigma \) muon \( g-2 \) requirement can now be reconsidered. Furthermore, models within this class predict a range of other implications, including non-zero neutrino masses and various coannihilation scenarios for DM phenomenology. All these factors combine to provide compelling reasons to re-investigate the phenomenology of supersymmetric 4-2-2 models.

The paper is organized as follows: In Section 2, we discuss the experimental constraints and scanning procedure used to explore the parameter space. Section 3 presents the results for the fundamental parameter space and sparticle mass spectrum. The implications for dark matter and the correlation with muon $g-2$ are examined in Section 4, and section 5 summarizes our conclusions.

\section{Experimental Constraints and Scanning Procedure}
\label{sec:scan}

The parameter space of the models in this class is spanned by the following parameters defined at $\mgut$: 

\begin{equation}
\begin{array}{lcl}
0 \leq & m_{L_{1,2}} & \leq 5 ~{\rm TeV} \\
0 \leq & m_{L_{3}} & \leq 10 ~{\rm TeV} \\
0\leq & {\rm x}_{{\rm LR}} & \leq 3 \\
0 \leq & M_{2L} & \leq 5 ~{\rm TeV} \\
-3 \leq & M_{3} & \leq 5 ~{\rm TeV} \\
-3 \leq & {\rm y}_{{\rm LR}} & \leq 3 \\
-3 \leq & A_{0}/{\rm max}(m_{L_{3}},m_{R_{3}}) & \leq 3 \\
35 \leq & \tan\beta & \leq 60 \\
0 \leq & {\rm x}_{{\rm d}} & \leq 3 \\
-1 \leq & {\rm x}_{{\rm u}} & \leq 2~.
\end{array}
\label{eq:paramSpacePSLR}
\end{equation}
$m_{L_{1,2}}$ and $m_{L_{3}}$ determine the SSB mass terms for the left-handed scalar matter fields of the first two-family and the third family respectively. With split families there is an additional free parameter in our scans. The different ranges set for these masses are based on the fact that the muon $g-2$ solutions need the first two-family sparticles light, while the 125 GeV Higgs boson mass requires heavy sparticles from the third-family.  With C-parity broken at $\mgut$, the soft scalar for the right and left scalar are different, i.e. $m_{L_{1,2}} \neq m_{R_{1,2}}$ and $m_{L_{3}} \neq m_{R_{3}}$. We parametrise this LR breaking with $x_{LR}$, namely $m_{R} \equiv x_{LR}m_{L}$, where we suppress the family indices since we assume the same $x_{LR}$ for all the families. Similarly, $M_{2L}$ and $M_{3}$ define the SSB gaugino masses associated with the $SU(2)_{L}$ and $SU(3)_{c}$ gauge groups. The MSSM hypercharge generator is given by \cite{Pati:1974yy} 
\begin{equation}
Y = \sqrt{\dfrac{3}{5}}I_{3R} + \sqrt{\dfrac{2}{5}}(B-L)\Rightarrow M_{1} = \dfrac{3}{5}M_{2R} + \dfrac{2}{5}M_{3},
\label{eq:LRgag}
\end{equation}
where $M_{2R}$ is the SSB gaugino mass associated with $SU(2)_{R}$, and in the LR symmetric models $M_{2R}=M_{2L}$. In our model, since, this symmetry does not hold, in our scans, we employ $y_{LR}$ as a measure of LR symmetry breaking in the gaugino sector by defining $M_{2R} \equiv y_{LR}M_{2L}$.

The set of free parameters also includes the trilinear scalar coupling ($A_{0}$) and $\tan\beta \equiv v_{u}/v_{d}$, where $v_{u,d}$ are the vacuum expectation values (VEVs) of the MSSM Higgs fields $H_{u}$ and $H_{d}$, respectively. Note that instead of varying $A_{0}$ directly, we control it by its ratio to the SSM scalar masses of the third family. We use the heaviest mass in the left- and right-handed sector and put an upper bound on this ratio to three in order to avoid the color/charge breaking minima of the scalar potential \cite{Ellwanger:1999bv,Camargo-Molina:2013sta}. We also define $x_{u}$ and $x_{d}$ to parametrise the non-universal SSB masses for the MSSM Higgs fields defined in our scans as $m_{H_{u}}^{2} = sgn(x_{u})x_{u}^{2}m_{L_{3}}^{2}$ and $m_{H_{d}}^{2} = sgn(x_{d})x_{d}^{2}m_{L_{3}}^{2}$. In addition to these parameters listed in Eq.(\ref{eq:paramSpacePSLR}), we set the $\mu > 0$ and the top quark mass to its central value ($m_{t}=173.3$ GeV \cite{CDF:2009pxd,ATLAS:2021urs}). Even though the SUSY mass spectrum is not very sensitive to variations in the top-quark mass, the SM-like Higgs boson mass can vary with it by $1-2$ GeV \cite{AdeelAjaib:2013dnf,Gogoladze:2014hca}.

To systematically explore this parameter space, we use random scans, employing SPheno \cite{Porod:2003um,Porod:2011nf}, a spectrum generator that calculates the SUSY masses, decay widths, and branching ratios, and micrOMEGAs \cite{Belanger:2018ccd} to calculate the dark matter (DM) observables, which are both generated by SARAH \cite{Staub:2008uz,Staub:2015iza}. After generating the data, we apply some constraints. Firstly, we only accept solutions which are compatible with the radiative electroweak symmetry breaking (REWSB) and a neutralino dark matter candidate. Next, collider constraints, especially those from the LHC, play a pivotal role. The absence of directly observed supersymmetric particles at the LHC places important lower bounds on their masses, especially for the supersymmetric particles carrying $SU(3)_c$ color charges, which exclude masses below about 2 TeV for the first two-family squarks and gluino \cite{ATLAS:2023afl,ATLAS:2020syg}. Additionally, rare decays of $B-$mesons \cite{CMS:2020rox,Belle-II:2022hys,HFLAV:2022pwe} provide stringent constraints on the parameter space, especially in scenarios with enhanced flavor violation. We also apply the latest measurements on the DM relic density by the Planck satellite \cite{Planck:2018nkj} to identify the compatible DM solutions. While the LSP solutions with larger relic density can be problematic, we also show the solutions which lead to a LSP neutralino relic density lower than the Planck measurements, since they can be interesting in models with an additional dark matter candidate such as an axion. The experimental constraints employed in our analyses are summarized below:

\begin{equation}
\setstretch{1.8}
\begin{array}{c}
m_h  = 123-127~{\rm GeV}\\
m_{\tilde{g}} \geq 2.1~{\rm TeV}\\
1.95\times 10^{-9} \leq{\rm BR}(B_s \rightarrow \mu^+ \mu^-) \leq 3.43 \times10^{-9} \;(2\sigma) \\
2.99 \times 10^{-4} \leq  {\rm BR}(B \rightarrow X_{s} \gamma)  \leq 3.87 \times 10^{-4} \; (2\sigma) \\
16.5 \times 10^{-10} \leq \Delta a_{\mu} \leq 24.54\times 10^{-10}\; (1\sigma) \\
0.114 \leq \Omega_{{\rm CDM}}h^{2} \leq 0.126~.
\label{constraints}
\end{array}
\end{equation}

In addition to these constraints, the models in this class predict Yukawa unification (YU) at $\mgut$ in their minimal construction \cite{Ananthanarayan:1991xp, Ananthanarayan:1992cd}. Even though it is imposed at the GUT scale, they have a strong impact on the low scale implications because of the need for the threshold corrections to the Yukawa couplings \cite{Hall:1993gn, Gogoladze:2010fu, Elor:2012ig}, which usually leads to heavy mass spectrum and disfavor large contributions to muon $g-2$. However, YU can be maintained only for the third-family Yukawa couplings, while it fails the consistent fermion masses for the first two-family. This inconsistency can be corrected by extending the Higgs sector and/or the GUT gauge group with a flavor symmetry \cite{Babu:2014sga}. If one extends the Higgs potential by involving the Higgs fields from ($1,2,2$), ($15,2,2$), ($15,1,3$) and ($15,1,1$) representations of $\PS$, non-zero VEVs in ($15,1,3$) direction can correct the mass relations among the first two-family fermions, while modifying the third-family yukawa couplings as follows (for details, see \cite{Gomez:2002tj,Dar:2011sj}):

\begin{equation}
y_{t}:y_{b}:y_{\tau} = (1+C):(1-C):(1+3C)~,
\label{eq:CQYU}
\end{equation}

In general Eq.(\ref{eq:CQYU}) can mean a complete breaking of YU, but one can define the Quasi YU by limiting the deviation from YU as $|C|\leq 0.2$.

Considering also the unification features of the $4-2-2$ class, we follow several scenarios, whose parameter spaces together can form the whole parameter space which is defined by the parameters lilsted in Eq.(\ref{eq:paramSpacePSLR}). We list these scenarios as follos:

\begin{enumerate}

\item $\PS$ with split families supplemented with the QYU condition (s422-QYU) \cite{Shafi:2023sqa} described as:

\begin{equation*}
m_{L_{1,2}} \neq m_{L_{3}}~; \hspace{0.3cm} x_{LR} \neq 1~{\rm and}~y_{LR} \neq 1~;\hspace{0.3cm} tan\beta \geq 35
\end{equation*}  
and constrained by the QYU condition defined as $|C|\leq 0.2$ through Eq.(\ref{eq:CQYU}).

\item $\PS$ with universal families (422-UNI): \cite{Gomez:2022qrb} described as 

\begin{equation*}
m_{L_{1,2}} = m_{L_{3}}~; \hspace{0.3cm} x_{LR} \neq 1~{\rm and}~y_{LR} \neq 1
\end{equation*}

\item Non-Universal gaugino mass models with split families (sNUGM) \cite{Shafi:2021jcg} described as

\begin{equation*}
m_{L_{1,2}} \neq m_{L_{3}}~; \hspace{0.3cm} x_{LR} = 1~{\rm and}~y_{LR} \neq 1
\end{equation*}

\end{enumerate}

\section{Fundamental Parameter Space of Muon $g-2$ and Mass Spectrum}
\label{sec:Spectrum}
\begin{figure}[ht!]
\centering
\subfigure{\includegraphics[scale=0.4]{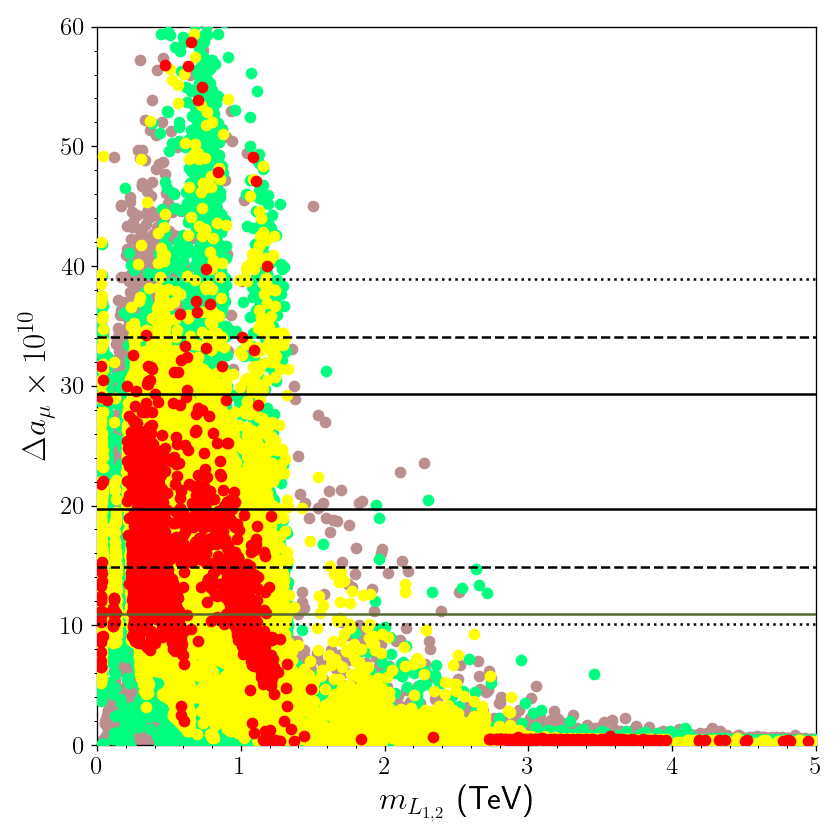}}%
\subfigure{\includegraphics[scale=0.4]{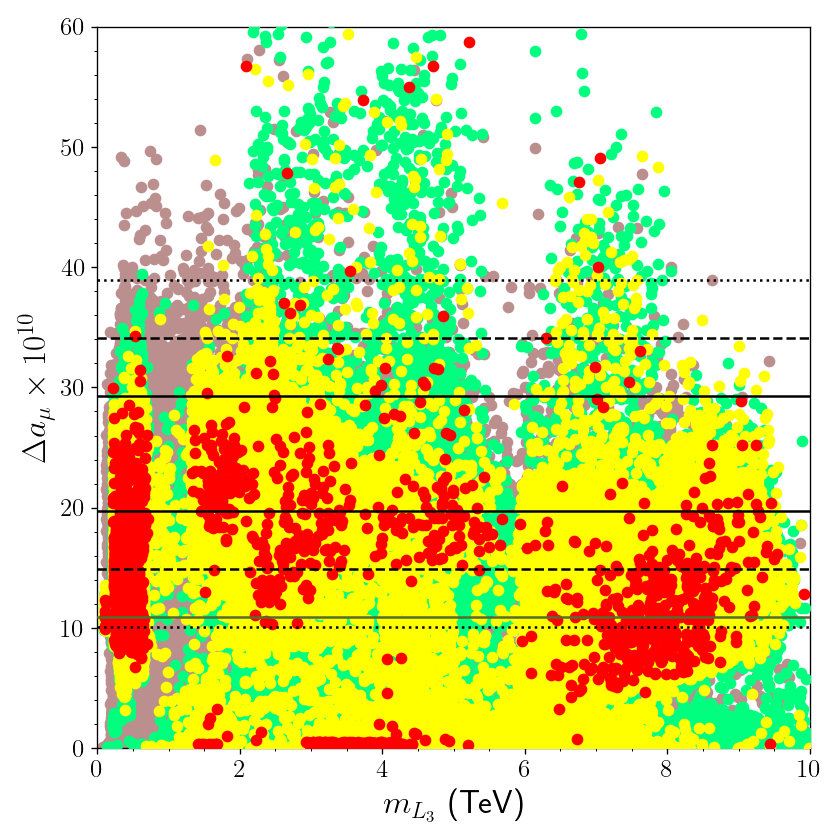}}
\subfigure{\includegraphics[scale=0.4]{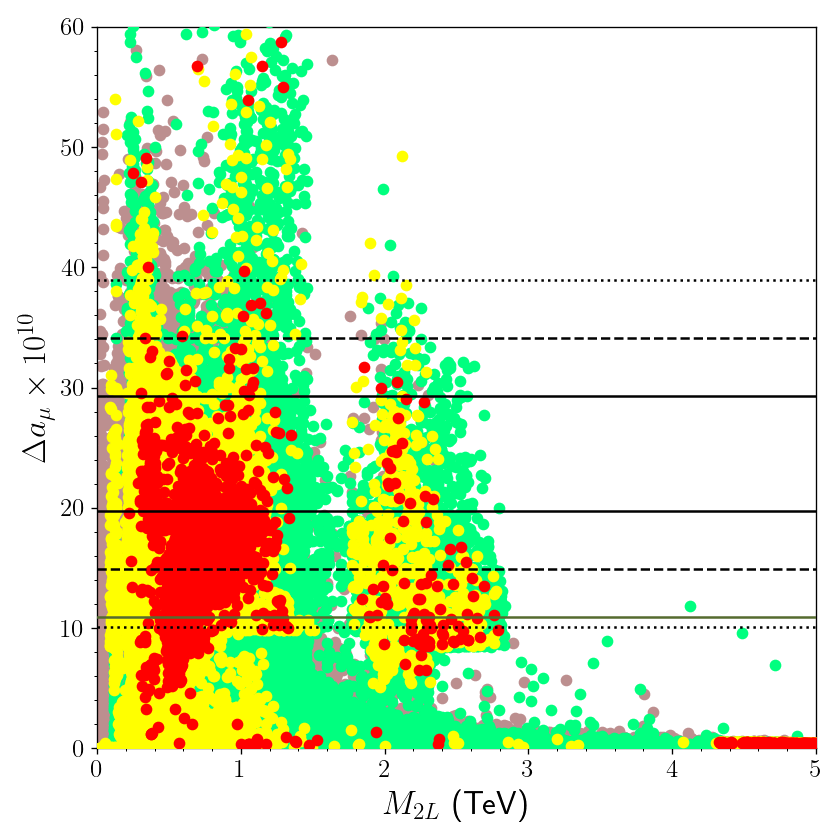}}%
\subfigure{\includegraphics[scale=0.4]{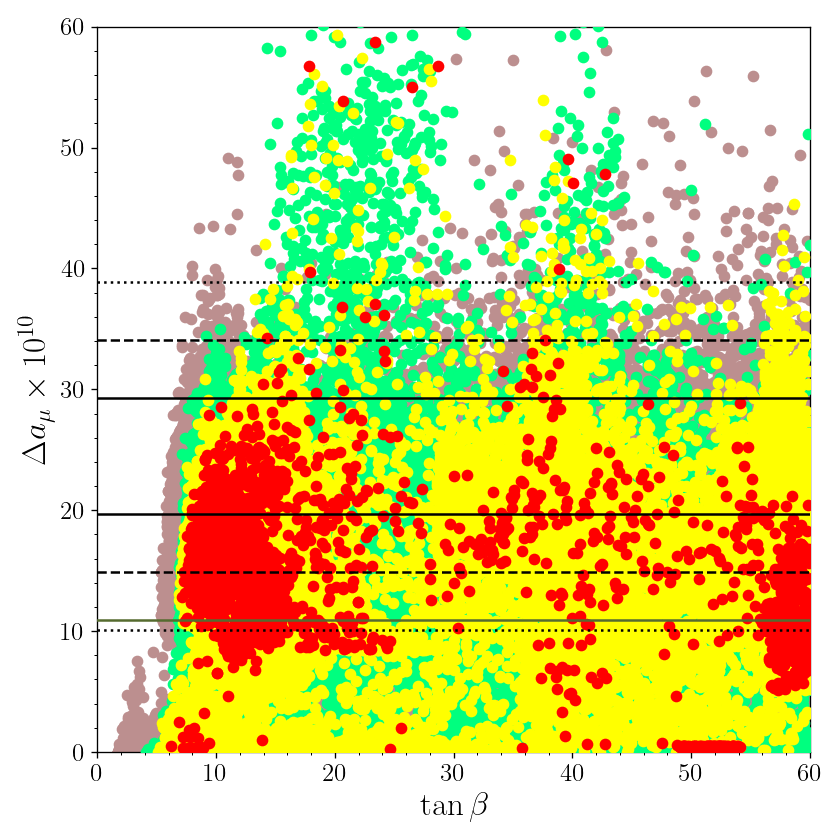}}
\caption{Plots for the muon $g-2$ results in terms of the SSB scalar masses (top), $M_{2L}$ and $\tan\beta$. All points are compatible with the REWSB and LSP neutralino condition. The green points satisfy the mass bounds and constraints from rare $B-$meson decays. The red points form a subset of green and they are compatible with the Planck measurements on relic density of LSP neutralino within $5\sigma$. The yellow points are another subset of green and they indicate the LSP neutralino solutions with relic density lower than the current Planck bound. The measurements of muon $g-2$ are displayed by the horizontal solid ($1\sigma$), dashed ($2\sigma$) and dotted ($3\sigma$) lines. The darkolive solid line indicates the deviation of BMW results from the FNAL measurement.}
\label{fig:damugutmasses}
\end{figure}

We begin with the updates in the fundamental parameter space by displaying the SSB masses at $\mgut$ for the scalars (top), $SU(2)_{L}$ gaugino (bottom-left) and $\tan\beta$ (top-right) in Figure \ref{fig:damugutmasses}. All points are compatible with the REWSB and LSP neutralino conditions. The green points satisfy the mass bounds and constraints from rare $B-$meson decays. The red points form a subset of green and they are compatible with the Planck measurements on the LSP relic density within $5\sigma$. The yellow points are another subset of green and they indicate the LSP neutralino solutions with relic density lower than the current Planck bound. The muon $g-2$ measurements are displayed by the horizontal solid ($1\sigma$), dashed ($2\sigma$) and dotted ($3\sigma$) lines. The most flexible model mentioned in the previous section is NUGM and the distributions on these parameters more or less coincide with the previous results except for some extensions in the range covered. The main impact from  muon $g-2$ solutions can be observed in the SSB masses of the first two families (top-left) and $SU(2)_{L}$ gaugino (bottom-left). The previous analyses used to bound $m_{L_{1,2}}$ at about 1 TeV from above, while its current measurements can be fulfilled for $m_{L_{1,2}} \lesssim 1.5$ TeV within $1\sigma$. The lower relic density solutions within $2\sigma$ of the muon $g-2$ results can move this bound a little further to about 1.6 TeV. Even though the bound on $M_{2L}$ from the muon $g-2$ results is not very strong, this parameter was previously bounded from above by the Planck measurements at about 1 TeV. The current analyses show that the solutions simultaneously compatible within $1\sigma$ of the muon $g-2$ and Planck measurements on the relic density of LSP can be realized in the region with $M_{2L}\lesssim 2.5$ TeV. As a result of the previous analyses, the predicted muon $g-2$ does not yield a direct impact on the third-family SSB mass and $\tan\beta$, which is displayed in the right panels of Figure \ref{fig:damugutmasses}. Even though large $\tan\beta$ is favored by the large supersymmetric contributions to muon $g-2$, this class of models can fit the muon $g-2$ solutions for $\tan\beta \gtrsim 8$.

\begin{figure}[ht!]
\centering
\subfigure{\includegraphics[scale=0.4]{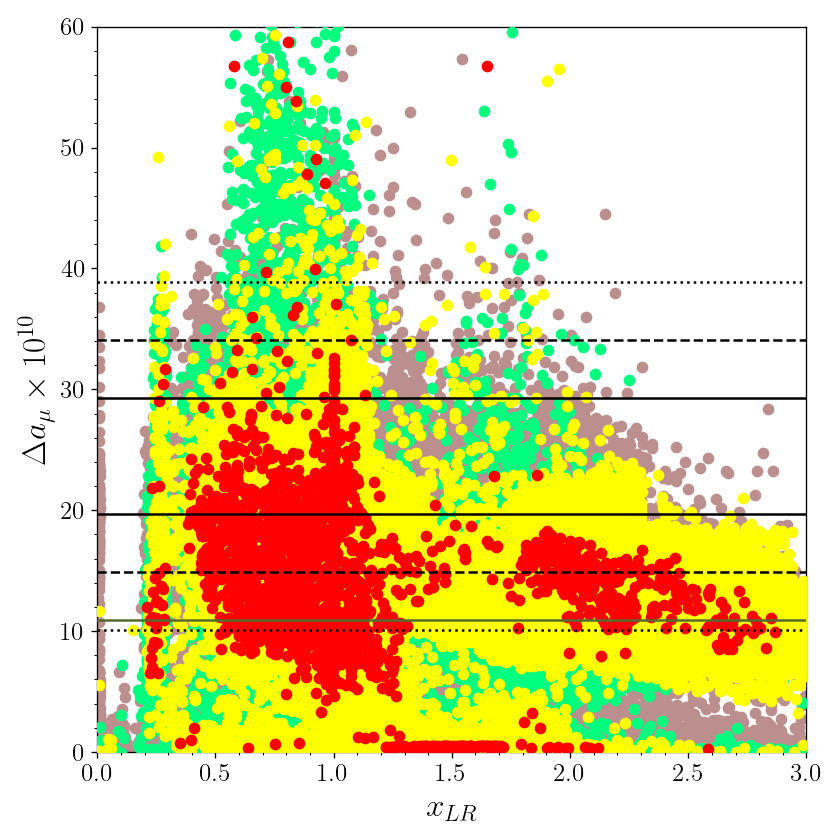}}%
\subfigure{\includegraphics[scale=0.4]{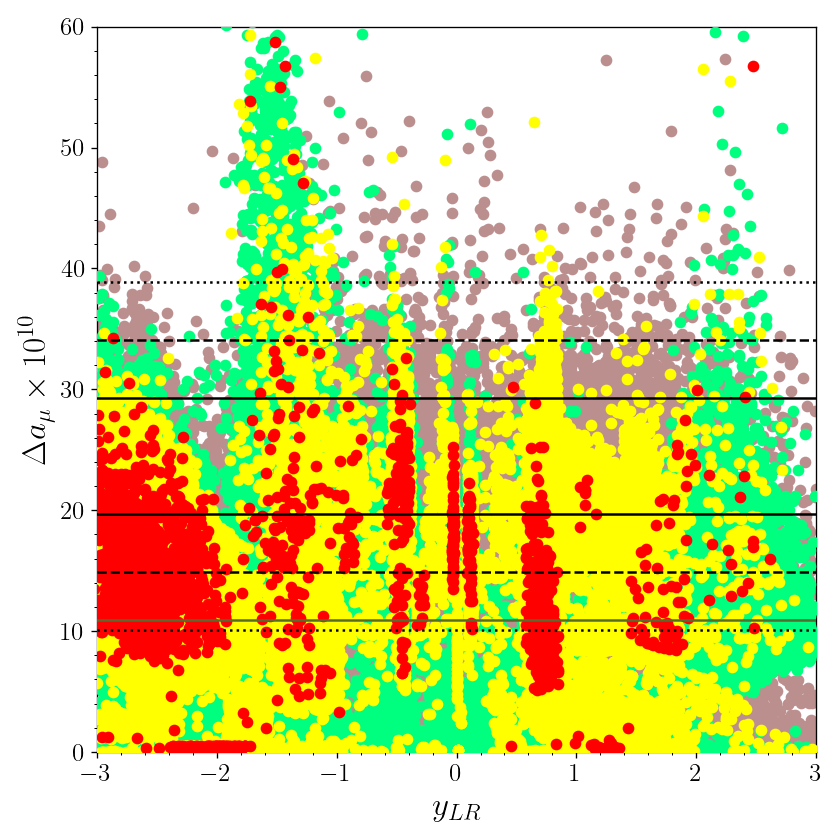}}
\caption{The muon $g-2$ results in correlation with the LR symemtry breaking parameters in the SSB scalar (left) and gaugino (right) sectors. The color coding is the same as in Figure \ref{fig:damugutmasses}.}
\label{fig:LRpars}
\end{figure}

\begin{figure}[ht!]
\centering
\subfigure{\includegraphics[scale=0.4]{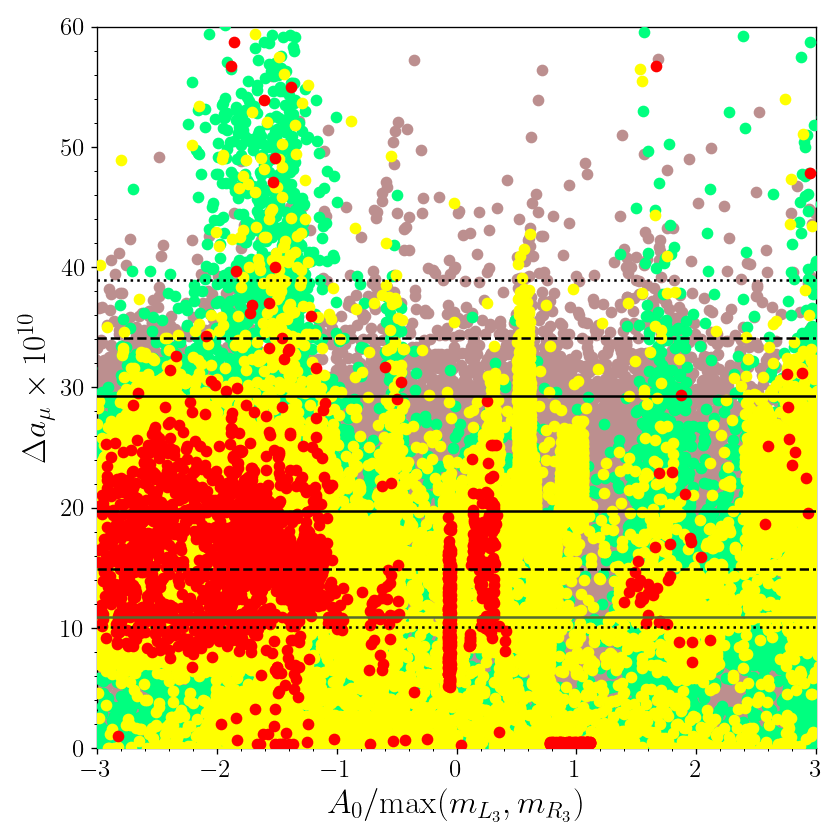}}%
\subfigure{\includegraphics[scale=0.4]{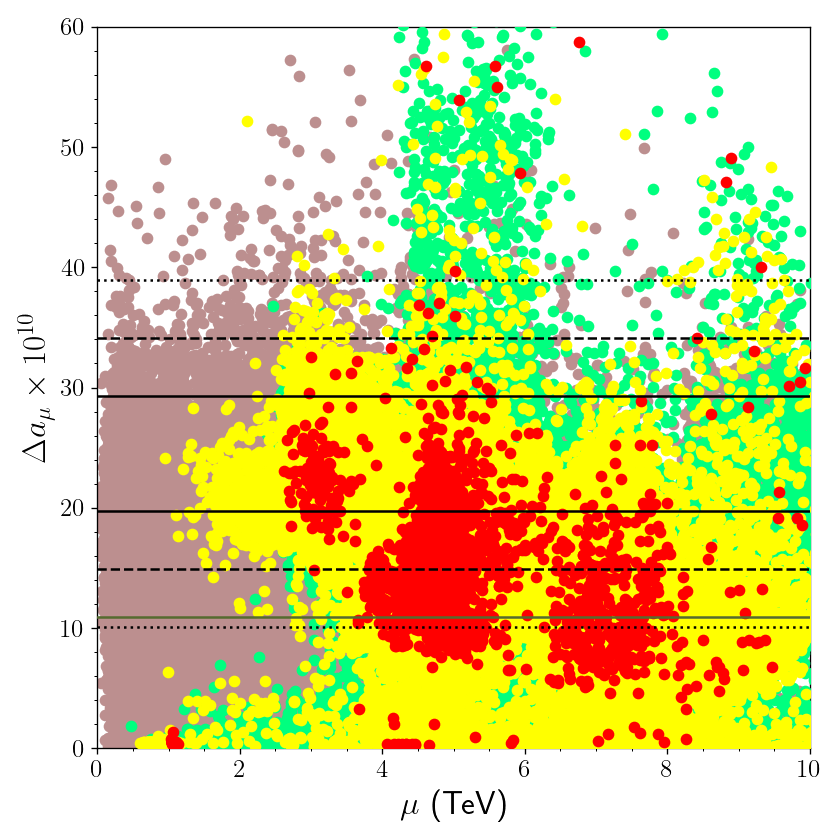}}
\caption{The plots in the $\Delta a_{\mu}-A_{0}/{\rm max}(m_{L_{3}},m_{R_{3}})$ and $\Delta a_{\mu}-\mu$ planes. The color coding is the same as in Figure \ref{fig:damugutmasses}.}
\label{fig:A0mu}
\end{figure}

\begin{figure}[ht!]
\centering
\subfigure{\includegraphics[scale=0.4]{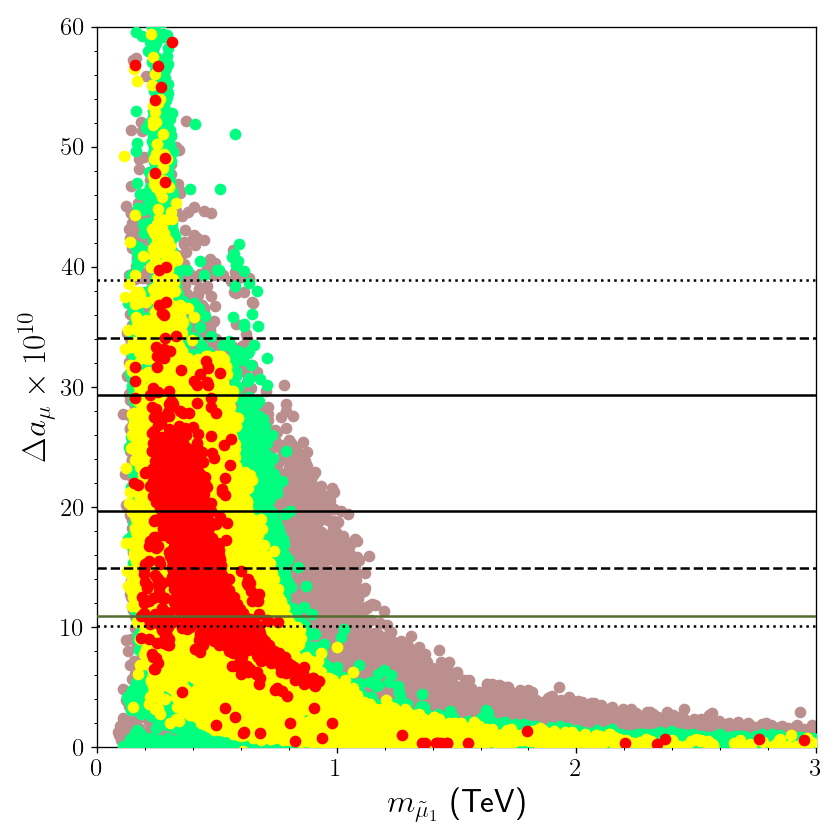}}%
\subfigure{\includegraphics[scale=0.4]{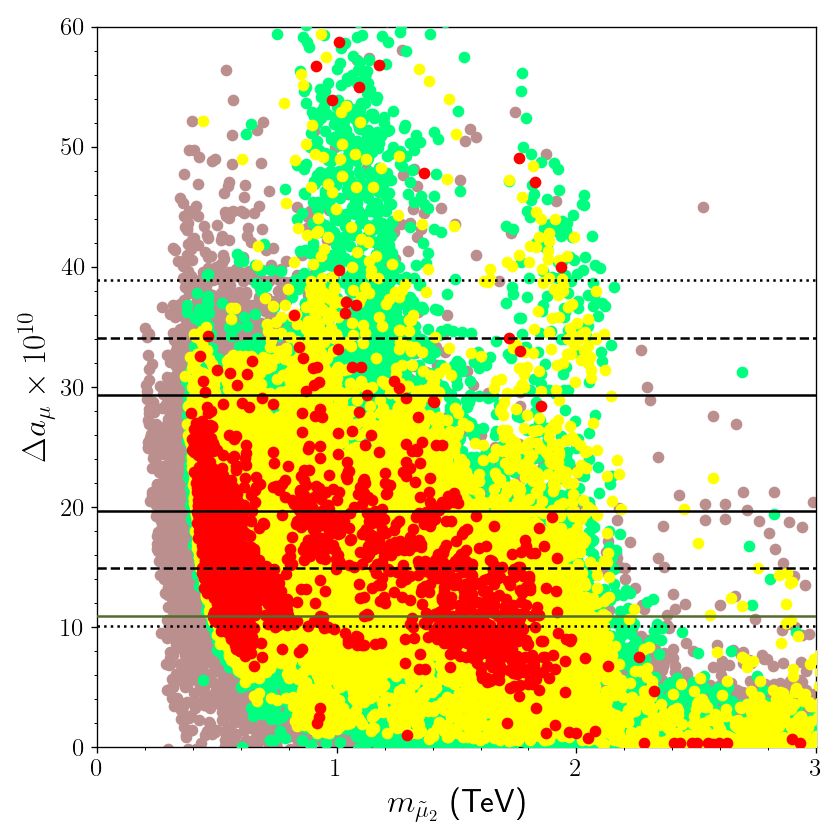}}
\subfigure{\includegraphics[scale=0.4]{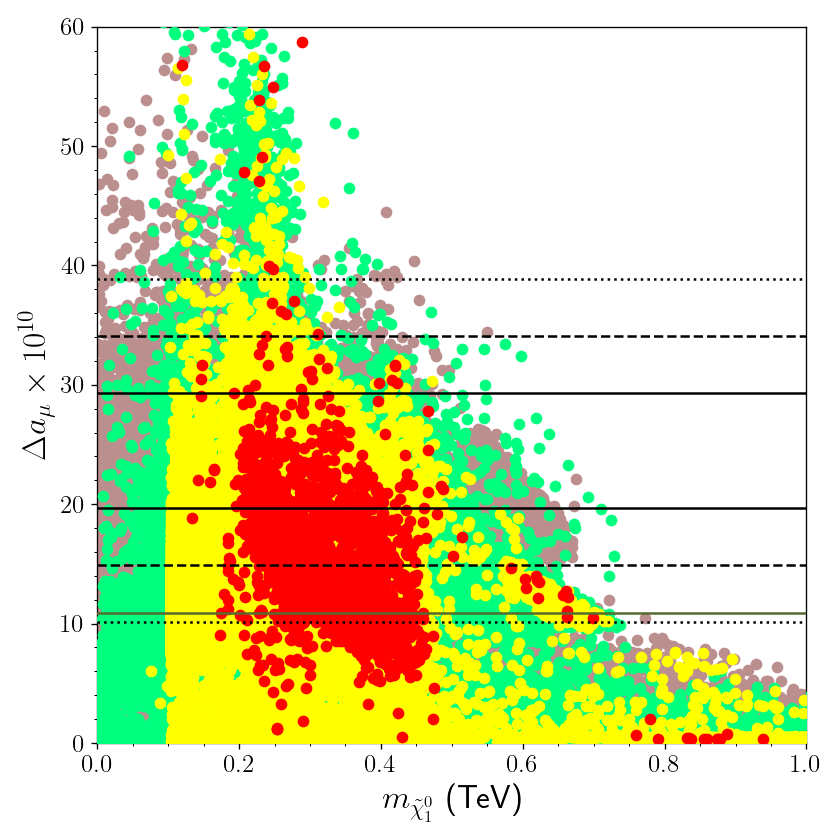}}%
\subfigure{\includegraphics[scale=0.4]{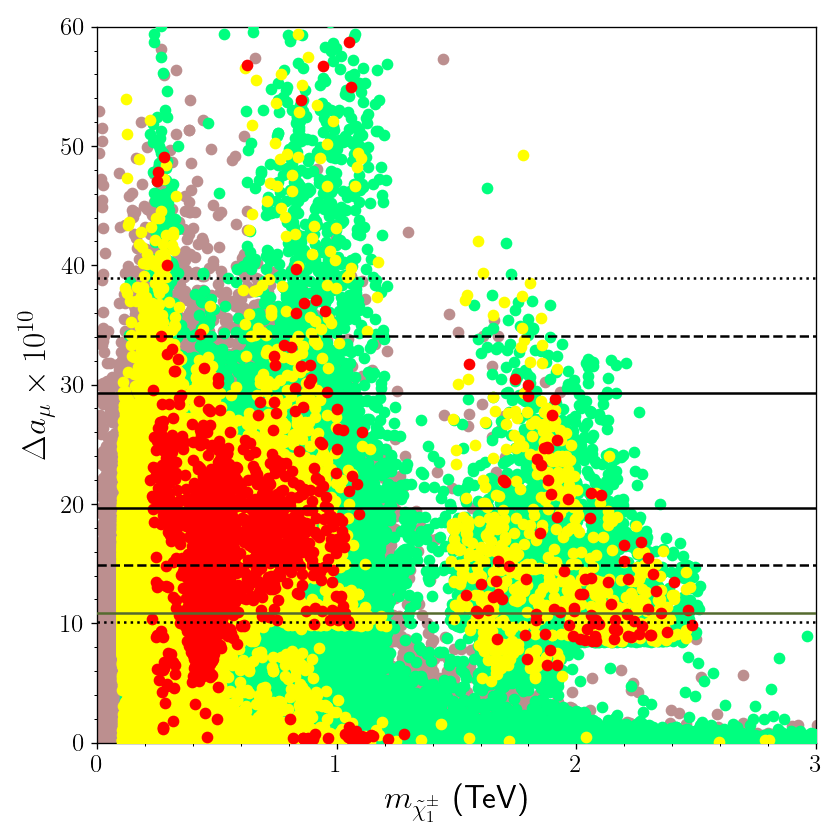}}
\caption{The plots for the mass spectrum in the $\Delta a_{\mu}-m_{\tilde{\mu}_{1}}$, $\Delta a_{\mu}-m_{\tilde{\mu}_{2}}$, $\Delta a_{\mu}-m_{\tilde{\chi}_{1}^{0}}$ and $\Delta a_{\mu}-m_{\tilde{\chi}_{1}^{\pm}}$ planes. The color coding is the same as in Figure \ref{fig:damugutmasses}.}
\label{fig:damulowmasses}
\end{figure}

\begin{figure}[ht!]
\centering
\subfigure[422-UNI]{\includegraphics[scale=0.4]{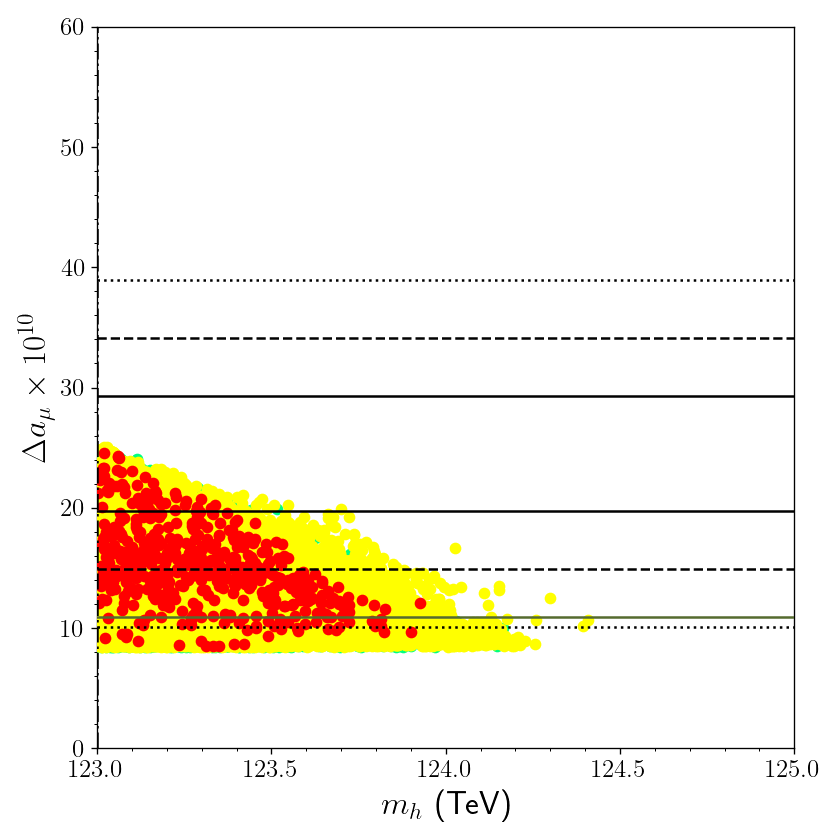}}%
\subfigure[s422-QYU]{\includegraphics[scale=0.4]{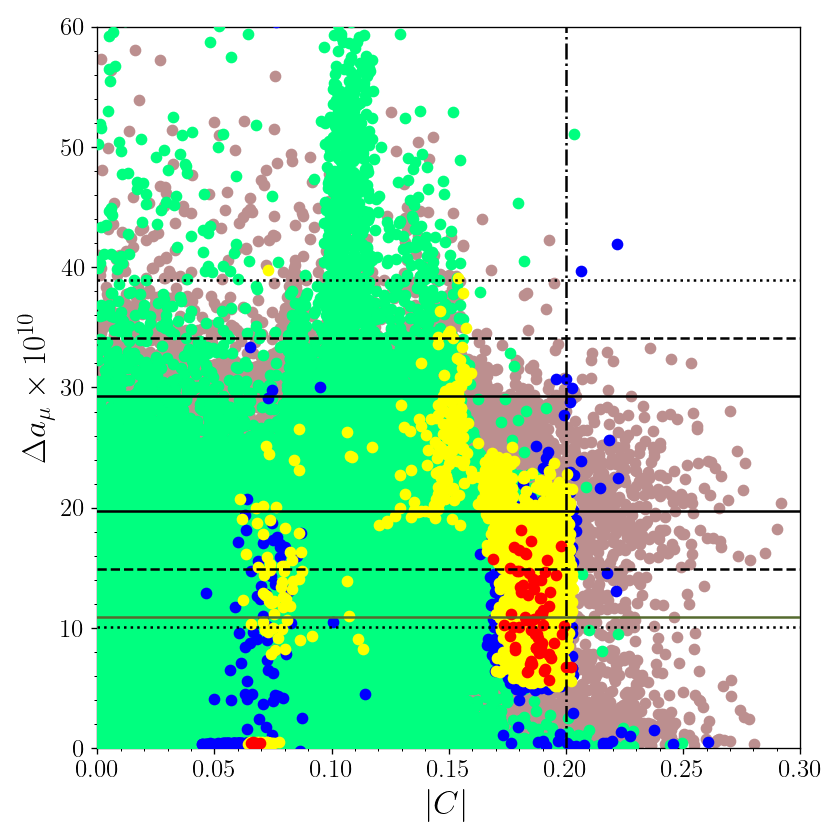}}
\caption{The SM-like Higgs boson mass in 422-UNI models (left) and the QYU impact on the muon $g-2$ solutions in s422-QYU models (right). The color coding is the same as in Figure \ref{fig:damugutmasses}. In the left plane the blue points indicate QYU solutions and, red and yellow points form distinct subsets of blue with the same meaning as in Figure \ref{fig:damugutmasses}. The vertical line in the right-plane shows the QYU limit on $|C|$ at $|C| = 0.2$.}
\label{fig:uniqyu}
\end{figure}

\begin{figure}[ht!]
\centering
\subfigure{\includegraphics[scale=0.4]{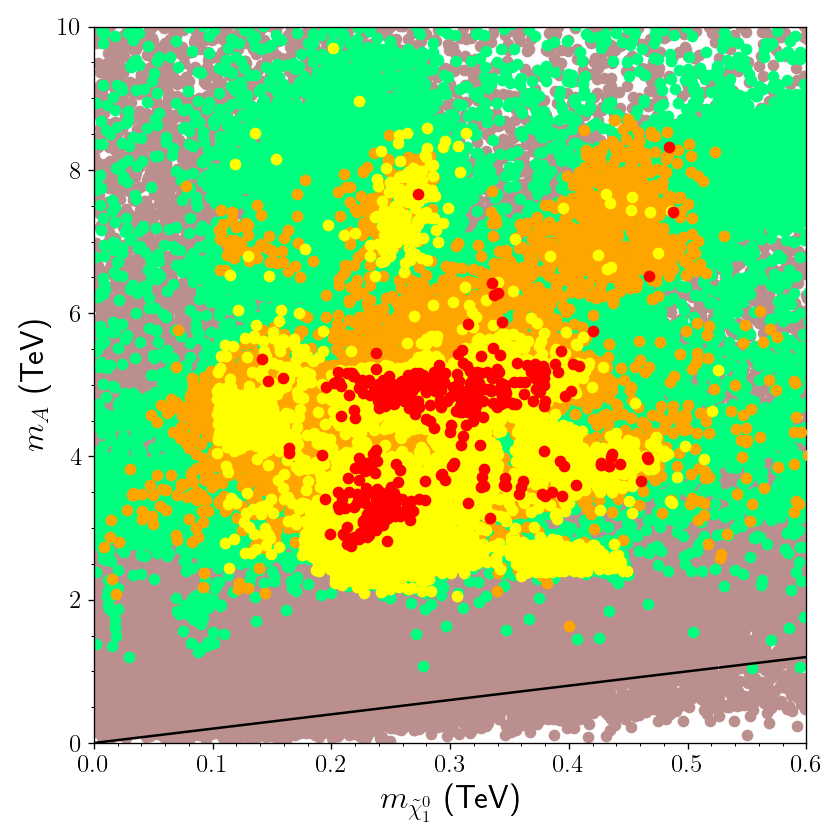}}%
\subfigure{\includegraphics[scale=0.4]{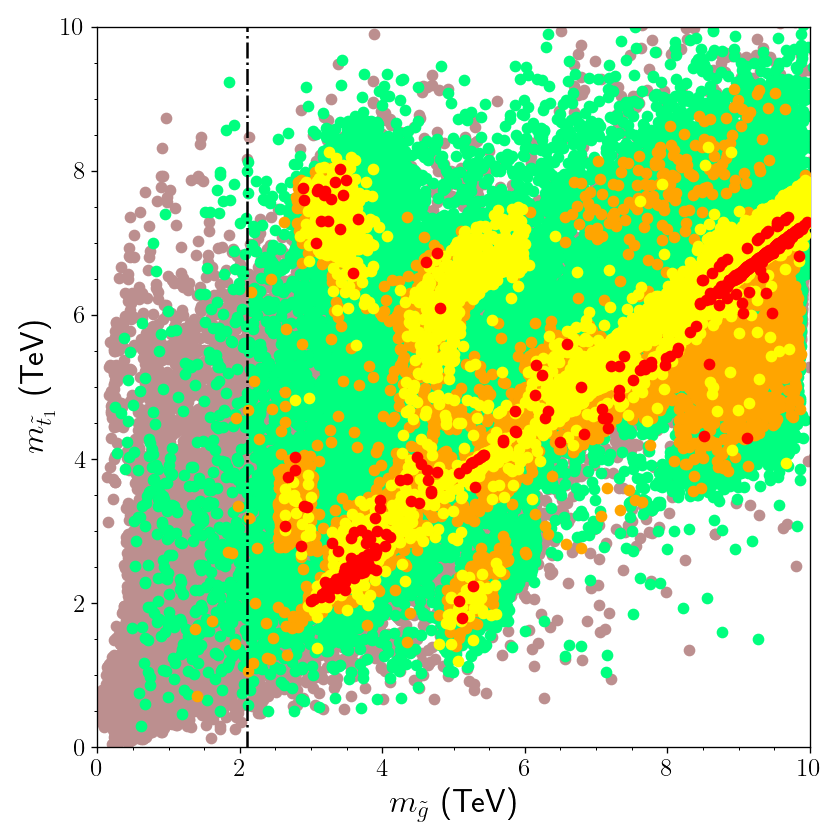}}
\caption{Plots in the $m_{A}-m_{\tilde{\chi}_{1}^{0}}$ and $m_{\tilde{t}_{1}}-m_{\tilde{g}}$ planes. All points are compatible with the REWSB and LSP neutralino conditions. Green points are consistent with the mass bounds and constraints from rare $B-$meson decays. The orange points form a subset of green and they display the solutions compatible with muon $g-2$ measurements within $1\sigma$. Red and yellow points are independent subsets of orange and they indicate the compatible DM (red) and low relic density LSP neutralino solutions (yellow) as those in Figure \ref{fig:damugutmasses}. The gluino mass bound is not applied in the right plane, which is shown by the vertical line.}
\label{fig:exclusionmass}
\end{figure}

We display the muon $g-2$ results with respect to the LR symmetry breaking parameters in Figure {\ref{fig:LRpars}}, with the color coding the same as in Figure \ref{fig:damugutmasses}. Even though it is possible to realize muon $g-2$ solutions compatible with the various constraints including the DM relic density within the whole ranges of these parameters, most of these solutions (red) are concentrated in the region with $x_{LR} \lesssim 2$.

We can conclude the discussion of the fundamental parameter space with the trilinear scalar coupling, which is displayed together with the MSSM $\mu-$term in Figure \ref{fig:A0mu}. The color coding is the same as in Figure \ref{fig:damugutmasses}. As can easily be seen, most of the solutions are concentrated in the negative $A_{0}$ region, but it is also possible to fit the muon $g-2$ solutions in the positive $A_{0}$ region. The region with $A_{0} < 0$ is somewhat favored by the SM-like Higgs boson mass, but it is possible to realize a consistent Higgs boson mass in the $A_{0} > 0$ region in our models. We also display the range of $\mu-$term compatible with the measured muon $g-2$ within $1\sigma$ and the other constraints. The relatively large muon $g-2$ contributions from the SUSY sector necessitate a large $\mu-$term, and, as seen from the $\Delta a_{\mu}-\mu$ plane, the $\mu-$term is bounded at about 2 TeV from below if the muon $g-2$ discrepancy is resolved to within $1\sigma$. This directly leads to the result that the Higgsinos do not take part in the LSP composition due to their heavy masses.

The impact of muon $g-2$ on $m_{L_{1,2}}$ and $M_{2L}$ arises from the fact that the supersymmetric contributions to the former arise directly from the smuon-neutralino and sneutrino-chargino loops \cite{Fargnoli:2013zia}. We show the masses of the relevant particles in Figure \ref{fig:damulowmasses} with plots in the $\Delta a_{\mu}-m_{\tilde{\mu}_{1}}$, $\Delta a_{\mu}-m_{\tilde{\mu}_{2}}$, $\Delta a_{\mu}-m_{\tilde{\chi}_{1}^{0}}$ and $\Delta a_{\mu}-m_{\tilde{\chi}_{1}^{\pm}}$ planes. The color coding is the same as in Figure \ref{fig:damugutmasses}. The $\Delta a_{\mu}-m_{\tilde{\mu}_{1}}$ plane shows that the muon $g-2$ solutions within $1\sigma$ can bound the lightest smuon at about 800 GeV, while the relic density constraint can slightly lower this bound. Note that the previous bound for the lightest smuon was about 600 GeV. Similarly, the LSP neutralino can be as heavy as about 600 GeV and the lightest chargino mass is bounded to be lighter than about 2.2 TeV, which are similar to those obtained in our previous analyses. These bounds on $m_{\tilde{\mu}_{1}}$ arise directly from the muon $g-2$ requirement, while an extension in the heavier smuon mass ranges can also result from the extension of the parameter space. If the LR symmetry is assumed in the scalar sector ($x_{LR}=1$), the muon $g-2$ solutions yield the bound $m_{\tilde{\mu}_{2}} \lesssim 1.5$ TeV \cite{Shafi:2021jcg}. In the case of fully broken LR symmetry ($x_{LR},y_{LR} \neq 1$), the heavier smuon can be as heavy as about 2 TeV in the muon $g-2$ region within $1\sigma$. 

Other interesting results can be seen for more restricted models such as 422-UNI and s422-QYU which can be read from the plots in Figure \ref{fig:uniqyu}, with the color coding the same as in Figure \ref{fig:damugutmasses}. The vertical line in the right-plane shows the QYU limit on $|C|$ at $|C| = 0.2$. We display the SM-like Higgs boson mass in the subspace of the parameter space which corresponds to 422-UNI models, with universal SSB scalar masses for the families ($m_{L_{1,2}} = m_{L_{3}}$). The previous analyses showed that the SM-like Higgs boson mass can be barely 123 GeV \cite{Gomez:2022qrb} (acceptable within theoretical uncertainties) for the muon $g-2$ region within $2\sigma$. On the other hand, the new analyses reveal that the current muon $g-2$ measurements can be accommodated within $2\sigma$, with the SM-like Higgs boson mass found to be 124 ($\pm 2$) GeV . Note that a 124 GeV Higgs boson mass can be fit with its experimental value of 125.6 GeV by employing the Himalaya three-loop corrections \cite{Allanach:2001kg,Allanach:2014nba,Baer:2021tta}, or by varying the top-quark mass within $1-2\sigma$ \cite{AdeelAjaib:2013dnf}. Another interesting observation is to realize muon $g-2$ solutions within $1\sigma$ compatible with the QYU scenario discussed in Ref. \cite{Shafi:2023sqa}. Our current analyses optimized to identify the regions compatible with muon $g-2$ within $1\sigma$ also bring out such solutions which we did not observe in our previous analyses.

Finally, we display the masses of the CP-odd Higgs boson mass (left) and strongly interacting supersymmetric particles (right) in Figure \ref{fig:exclusionmass} with plots in the $m_{A}-m_{\tilde{\chi}_{1}^{0}}$ and $m_{\tilde{t}_{1}}-m_{\tilde{g}}$ planes. All points are compatible with REWSB and LSP neutralino as we mentioned earlier. Green points are consistent with the mass bounds and constraints from rare $B-$meson decays. The orange points form a subset of green and they display the solutions compatible with the muon $g-2$ measurements within $1\sigma$. The red and yellow points are independent subsets of orange and they indicate the compatible DM (red) and low relic density LSP neutralino solutions as those in Figure \ref{fig:damugutmasses}. The gluino mass bound is not applied in the right plane, which is shown by the vertical line. The CP-odd Higgs boson mass is bounded as $m_{A} \gtrsim 2$ TeV by the LSP relic density constraint, while the muon $g-2$ within $1\sigma$ allows it to be as light as about 1 TeV. This bound from the DM relic density is also compatible with those from the analyses for $A,H \rightarrow \tau\tau$ events \cite{ATLAS:2020zms,CMS:2022goy}. The stop and gluino can also be as light as about 1 TeV compatible with the large muon $g-2$ contributions, while the DM constraint also bounds them at about 2 TeV from below.

\section{Dark Matter Implications and Muon $g-2$}
\label{sec:DM}

\begin{figure}[t!]
\centering
\subfigure{\includegraphics[scale=0.4]{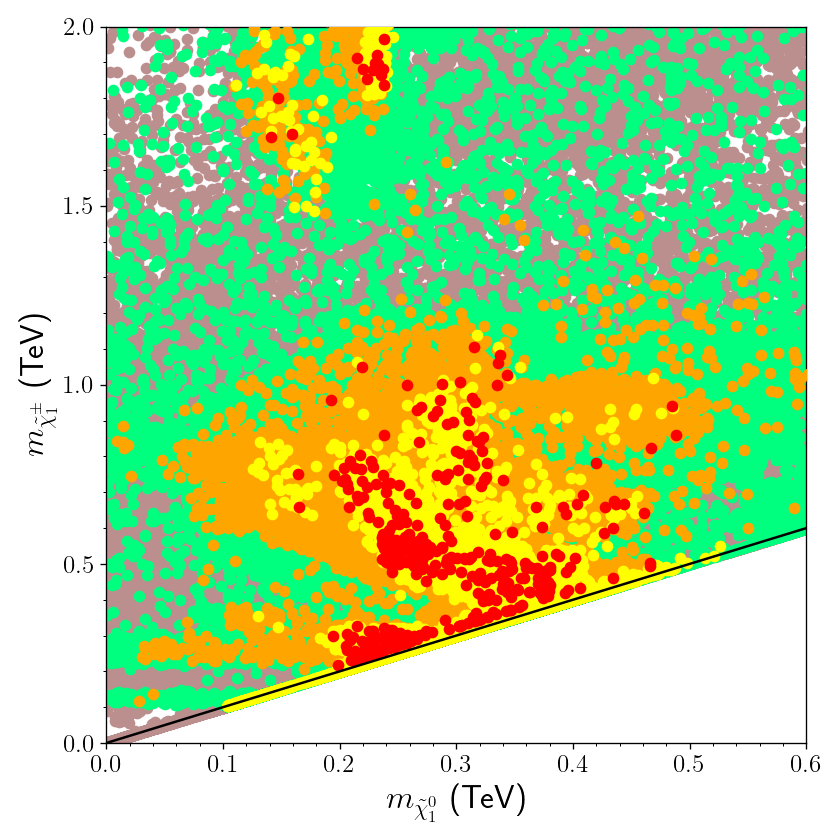}}%
\subfigure{\includegraphics[scale=0.4]{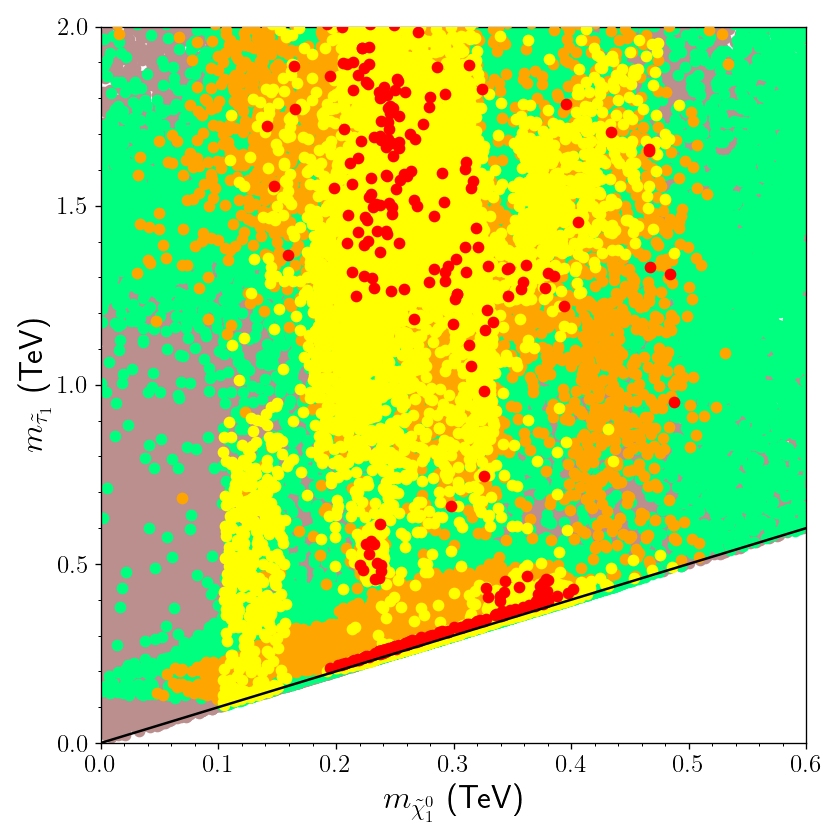}}
\caption{Chargino (left) and Stau (right) masses in correlation with the LSP neutralino mass. The color coding is the same as in Figure \ref{fig:exclusionmass}. The diagonal lines show the solutions in which the plotted particles are degenerate in mass.}
\label{fig:coans}
\end{figure}

\begin{figure}[t!]
\centering
\subfigure{\includegraphics[scale=0.4]{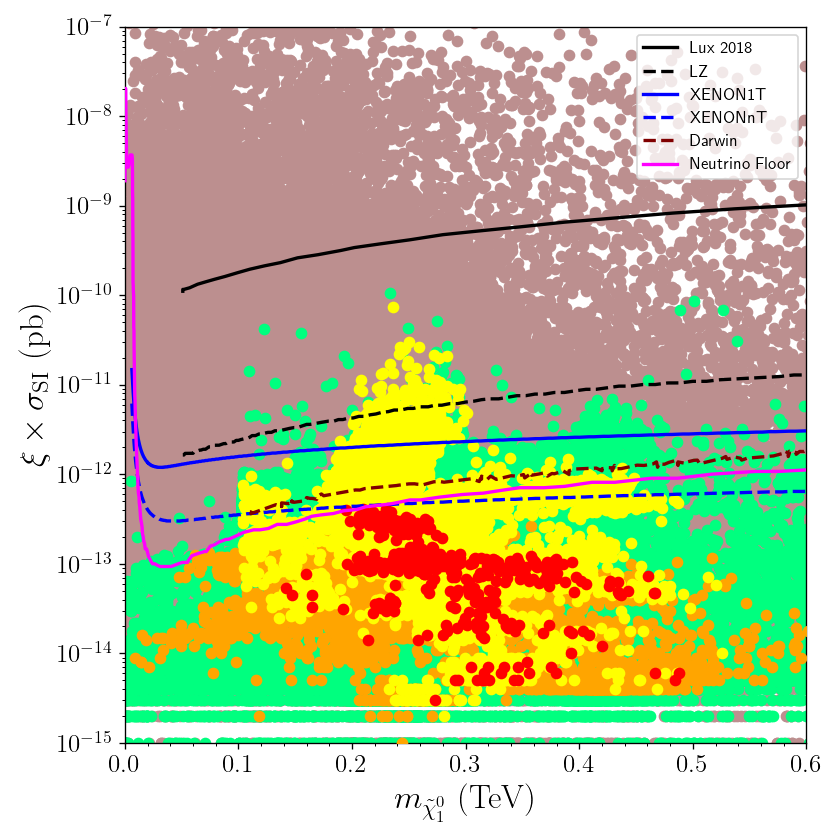}}%
\subfigure{\includegraphics[scale=0.4]{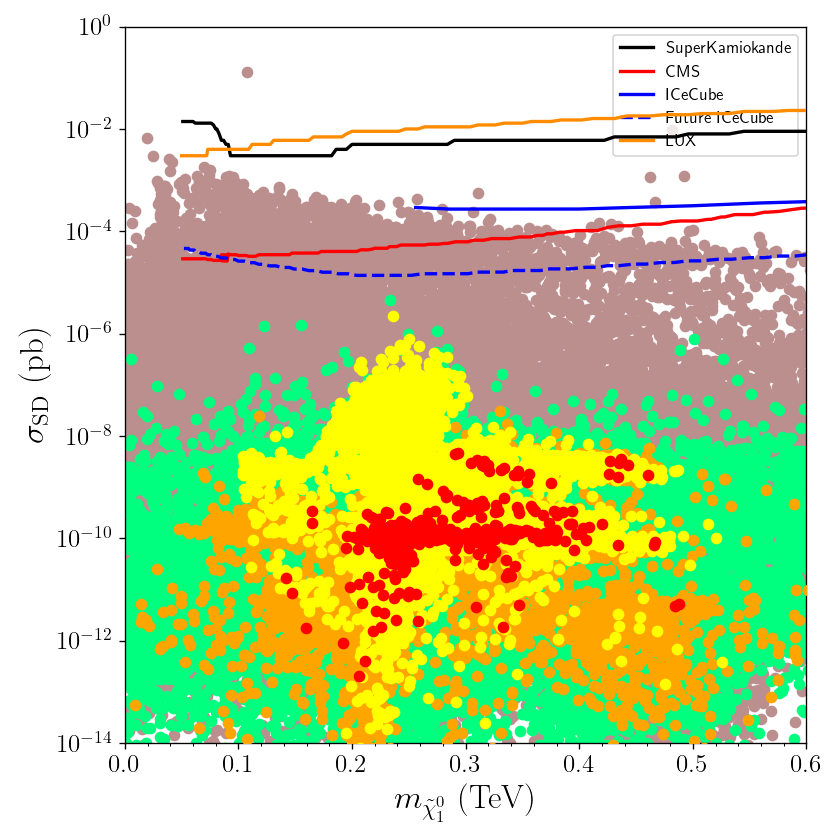}}
\caption{Spin-independent (left) and spin-dependent (right) scattering cross-sections of the LSP neutralino. The color coding is the same as in Figure \ref{fig:exclusionmass}. The curves represent the current and projected exclusion curves from several direct detection DM experiments, with the color coding given in each panel. The current excluded regions are represented by the solid curves, and the dashed curves display the projected experimental sensitivity.}
\label{fig:SISD}
\end{figure}

In this section we present our predictions for several coannihilation scenarios and DM observables which can be tested in the current and near future experiments. Figure \ref{fig:coans} displays the masses of the chargino (left) and stau (right) in correlation with the LSP neutralino mass. The color coding is the same as in Figure \ref{fig:exclusionmass}. The diagonal lines show the solutions in which the plotted particles are degenerate in mass. As seen from both planes, the DM solutions with a relic density compatible with the Planck measurements within $5\sigma$ (red) can be realized in the LSP mass range $200 \lesssim m_{\tilde{\chi}_{1}^{0}}\lesssim 600$ GeV. One can identify solutions in this region corresponding to the chargino-neutralino and stau-neutralino coannhilation channels. Even though the chargino-neutralino coannihilation scenario is one of the typical signs of the Wino-like and Higgsino-like LSP, in the $1\sigma - $muon $g-2$ region, the desired relic density solutions compatible with the Planck measurements (red) happen if the LSP is essentially the Bino, and the chargino-neutralino coannihilation solutions can be identified as $M_{\tilde{B}} \lesssim M_{\tilde{W}} \ll \mu$. In addition to the Bino-like LSP, it is also possible to realize Wino-like LSP solutions, but they usually lead to lower relic density for the LSP neutralino. Similarly, there are also solutions for the stau-neutralino coannihilation scenario. Although we only display the stau mass in Figure \ref{fig:coans}, the other sleptons (smuons and selectrons) also take part in coannihilation processes in this region, and some solutions predicting multiple coannihilation scenarios lead to a low LSP relic density. Even though these solutions are not compatible with the Planck measurements, they can be relevant in a setting with two or more DM candidates. 

Figure \ref{fig:SISD} displays our results for the spin-independent (left) and spin-dependent (right) scattering cross-sections in correlation with the LSP neutralino mass. The color coding matches that in Figure \ref{fig:exclusionmass}. The current and future exclusion curves from experiments such as LUX, LZ \cite{Akerib:2018lyp}, XENON \cite{Aprile:2020vtw} and DARWIN \cite{Aalbers:2016jon} for the spin-independent scattering cross-sections are also shown. The spin-dependent results are also confronted with the relevant experimental results as given in the legend \cite{Tanaka:2011uf,Khachatryan:2014rra,Abbasi:2009uz,Akerib:2016lao}. As mentioned above, the solutions with the desired relic density (red) lead to Bino-like LSP neutralino, and they typically lead to low spin-independent scattering cross-sections as depicted with the red points in the $\sigma_{{\rm SI}}-m_{\tilde{\chi}_{1}^{0}}$ plane. Most of these solutions lie below the neutrino floor, and more experimental data is necessary to test them. However, it is still possible to realize some solutions around the neutrino floor or slightly above it. These solutions are expected to be tested by the XENON collaboration and DARWIN. We also show the predictions for spin-dependent cross-section in the right panel of Figure \ref{fig:SISD} compared to the experimental limits which are listed in the legend with the corresponding color coding. While the experimental results are less sensitive in comparison with those from the spin-independent analyses, some solutions may be tested after future IceCube upgrades.

Finally, before concluding, we present two tables of benchmark points (BPs) which exemplify our findings. Table \ref{tab:muong2} displays five BPs which shed light on various coannihilation scenarios and their implications. All points adhere to the constraints imposed in our analyses and are selected to showcase the richness of the parameter space. Masses are given in GeV, and the DM scattering cross-sections are expressed in pb. We have highlighted the coannihilation candidates in red for clarity. Point 1 highlights a neutralino-chargino coannihilation scenario to achieve the observed relic density. Point 2 presents a neutralino-stau coannihilation scenario. The stau-neutralino coannihilation processes correspond to the third family SSB mass ($m_{L_{3}}$) less than about 500 GeV, the SM-like Higgs boson mass receives relatively small contributions from the third family, and it is accommodated in the spectrum at about 124 ($\pm 2$) GeV. Point 3 exemplifies a mixed coannihilation scenario involving neutralinos, charginos, sneutrinos, selectrons, smuons, and stau. Given the variety of coannihilation channels, the relic density is compatible with the current Planck bound through a combination of these processes. Point 4 shows a neutralino-selectron/smuon coannihilation scenario. This point also shows the SM like Higgs boson mass, which is close to its  experimentally measured value. We have depicted a solution exemplifying the Wino-like LSP neutralino compatible with the muon $g-2$ measurements. There are also several coannihilation scenarios for this case, which significantly lowers the LSP relic density.

Table $\ref{tab:QYU}$ focuses on benchmark points compatible with the current muon $g-2$ measurements within $2\sigma$ and QYU. Point 1 depicts a solution which can accommodate QYU in the region of muon $g-2$ within $1\sigma$. Such solutions can be classified under chargino-neutralino coannihilation, but the coannihilation processes significantly lower the LSP relic density. One can find solutions with the desired LSP relic density if the muon $g-2$ condition is relaxed to $2\sigma$ as shown in Point 2 of Table \ref{tab:QYU}. We depict a Higgsino-like LSP solutions with Point 3. Since large contributions to muon $g-2$ from the supersymmetric particles favor a large $\mu-$ term, the Higgsinos do not provide muon $g-2$ compatible solutions even within $3\sigma$, as seen from Point 3. Even though we display a QYU compatible solution here, Point 3 represents the Higgsino-like LSP solutions.

\begin{table}[h!]
\centering
\setstretch{1.5}
\scalebox{0.87}{
\begin{tabular}{|c|ccccc|}
\hline  & Point 1 & Point 2 & Point 3 & Point 4 & Point 5 \\ \hline
$m_{\tilde{L}_{1,2}}$ & 484.9 & 376.4 & 442.8 & 745.8 & 453.1 \\
$m_{\tilde{L}_{3}}$ & 2684 & 539.4 & 590.2 & 3948 & 2131 \\
$M_{1}$ & 1099 & 910.1 & 1008 & 651.1 & 1530 \\
$M_{2}$ & 638.1 & 667.4 & 610.9 & 1035 & 803.8 \\
$M_{3}$ & 3651 & 4956 & 4969 & 3732 & 4150 \\
$A_{0}/m_{\tilde{L}_{3}}$ & -1.2 & -2 & -1.6 & -1.5 & -0.668 \\
$\tan\beta$ & 36.8 & 11.1 & 13.1 & 18.6 & 59.5 \\ \hline
$x_{LR}$ & 0.752 & 0.54 & 0.668 & 0.744 & 0.898 \\
$y_{LR}$ & -0.946 & -2.7 & -2.7 & -1.4 & -0.269 \\
$m_{\tilde{R}_{1,2}}$ & 364.9 & 203.3 & 295.9 & 555.2 & 406.9 \\
$m_{\tilde{R}_{3}}$ & 2019 & 291.4 & 394.4 & 2939 & 1915 \\ \hline
$\mu$ & 4769 & 5132 & 5119 & 4721 & 4703 \\
$\Delta a_{\mu}\times 10^{10}$ & 21 & 20.6 & 16.4 & 24.4 & 20 \\ \hline
$m_{h}$ & 124.5 & 123.9 & 124 & 125.5 & 123.1 \\
$m_{H}$ & 3960 & 5228 & 5166 & 5060 & 2861 \\
$m_{A}$ & 3960 & 5228 & 5166 & 5060 & 2861 \\
$m_{H^{\pm}}$ & 3959 & 5225 & 5163 & 5059 & 2862 \\ \hline
$m_{\tilde{\chi}_{1}^{0}}$,$m_{\tilde{\chi}_{2}^{0}}$ & {\color {red} 466.3}, {\color{red} 494.1} & {\color{red} 369.1}, 487.9 & {\color{red} 412.8}, {\color{red}439.5} & {\color{red}268}, 841.5 & {\color{red} 630.9}, {\color{red}659.4} \\
$m_{\tilde{\chi}_{3}^{0}}$,$m_{\tilde{\chi}_{4}^{0}}$ & 4773, 4773 & 5165, 5165 & 5151, 5152 & 4739, 4739 & 4694, 4694 \\
$m_{\tilde{\chi}_{1}^{\pm}}$,$m_{\tilde{\chi}_{2}^{\pm}}$ & {\color{red} 494.3}, 4774 & 488.1, 5165 & {\color{red} 439.6}, 5152 & 841.5, 4740 & {\color{red}631}, 4694 \\ \hline
$m_{\tilde{g}}$ & 7388 & 9815 & 9847 & 7523 & 8317 \\
$m_{\tilde{u}_{1}}$,$m_{\tilde{u}_{2}}$ & 6251, 6263 & 8270, 8272 & 8301, 8302 & 6370, 6379 & 7037, 7040 \\
$m_{\tilde{t}_{1}}$,$m_{\tilde{t}_{2}}$ & 5434, 6106 & 7144, 7722 & 7187, 7750 & 5101, 6584 & 6229, 6517 \\ \hline
$m_{\tilde{d}_{1}}$,$m_{\tilde{d}_{2}}$ & 6260, 6263 & 8273, 8276 & 8302, 8307 & 6351, 6379 & 7039, 7040 \\
$m_{\tilde{b}_{1}}$,$m_{\tilde{b}_{2}}$ & 6092, 6227 & 7714, 8242 & 7741, 8261 & 6577, 6815 & 6492, 6635 \\ \hline
$m_{\tilde{\nu}_{e}}$,$m_{\tilde{\nu}_{\tau}}$ & 527.6, 2538 & 405.1, 538.6 & {\color{red} 444.6}, 566.4 & 999.7, 3923 & {\color{red}631}, 1964 \\
$m_{\tilde{e}_{1}}$,$m_{\tilde{e}_{2}}$ & 531.3, 640.1 & 411.6, 458.3 & {\color{red}450.8}, 525.2 & {\color{red}272}, 1003 & {\color{red} 631.9}, 732.9 \\
$m_{\tilde{\tau}_{1}}$,$m_{\tilde{\tau}_{2}}$ & 1653, 2545 & {\color{red} 383.2}, 601.5 & {\color{red}438.1}, 652.6 & 2658, 3924 & 1494, 2004 \\ \hline
$\sigma_{{\rm SI}}$ & $ 4.7 \times 10^{-14} $ & $ 9.4 \times 10^{-14} $ & $ 8.4 \times 10^{-14} $ & $ 4.7 \times 10^{-14} $ & $ 7.88 \times 10^{-13} $ \\
$\sigma_{{\rm SD}}$ & $ 8.31 \times 10^{-11} $ & $ 9.47 \times 10^{-11} $ & $ 1.04 \times 10^{-10} $ & $ 8.5 \times 10^{-11} $ & $ 2.41 \times 10^{-9} $ \\
$\Omega h^{2}$ & 0.125 & 0.124 & 0.117 & 0.123 & 0.029 \\ \hline
\end{tabular}}
\caption{Benchmark points that exemplify our findings compatible with the muon $g-2$ measurements within $1\sigma$. The points are chosen to be consistent with all the applied constraints including the Planck bound on dark matter except Point 5. The masses are given in GeV and the cross-sections in pb.}
\label{tab:muong2}
\end{table}

\begin{table}[h!]
\centering
\setstretch{1.5}
\scalebox{0.87}{
\begin{tabular}{|c|ccc|}
\hline  & Point 1 & Point 2 & Point 3 \\ \hline
$m_{\tilde{L}_{1,2}}$ & 419.7 & 461 & 4507 \\
$m_{\tilde{L}_{3}}$ & 2308 & 1926 & 4150 \\
$M_{1}$ & 1287 & 1331 & -6659 \\
$M_{2}$ & 728 & 758.5 & 4929 \\
$M_{3}$ & 3523 & 3628 & -2819 \\
$A_{0}/m_{\tilde{L}_{3}}$ & -0.65 & -0.675 & 1.1 \\
$\tan\beta$ & 58.6 & 58.7 & 51.3 \\ \hline
$x_{LR}$ & 0.922 & 1 & 1.3 \\
$y_{LR}$ & -0.279 & -0.264 & -1.9 \\
$m_{\tilde{R}_{1,2}}$ & 386.9 & 479.4 & 5686 \\
$m_{\tilde{R}_{3}}$ & 2128 & 2003 & 5236 \\ \hline
$\mu$ & 4269 & 4263 & 1088 \\
$\Delta a_{\mu}\times 10^{10}$ & 24.5 & 19.7 & 0.364 \\ \hline
$m_{h}$ & 123.4 & 123.1 & 123.3 \\
$m_{H}$ & 2660 & 2652 & 4924 \\
$m_{A}$ & 2660 & 2652 & 4924 \\
$m_{H^{\pm}}$ & 2661 & 2653 & 4925 \\ \hline
$m_{\tilde{\chi}_{1}^{0}}$,$m_{\tilde{\chi}_{2}^{0}}$ & {\color{red}551.9}, {\color{red}573.2} & {\color{red}571.2}, {\color{red}597.6} & {\color{red}1089}, {\color{red}1090} \\
$m_{\tilde{\chi}_{3}^{0}}$,$m_{\tilde{\chi}_{4}^{0}}$ & 4259, 4259 & 4253, 4254 & 3074, 4219 \\
$m_{\tilde{\chi}_{1}^{\pm}}$,$m_{\tilde{\chi}_{2}^{\pm}}$ & {\color{red}573.4}, 4260 & {\color{red}597.8}, 4254 & {\color{red}1090}, 4219 \\ \hline
$m_{\tilde{g}}$ & 7140 & 7336 & 6049 \\
$m_{\tilde{u}_{1}}$,$m_{\tilde{u}_{2}}$ & 6053, 6059 & 6226, 6228 & 7291, 7522 \\
$m_{\tilde{t}_{1}}$,$m_{\tilde{t}_{2}}$ & 5443, 5730 & 5574, 5775 & 5205, 5409 \\ \hline
$m_{\tilde{d}_{1}}$,$m_{\tilde{d}_{2}}$ & 6057, 6059 & 6228, 6230 & 7291, 7523 \\
$m_{\tilde{b}_{1}}$,$m_{\tilde{b}_{2}}$ & 5704, 5835 & 5746, 5957 & 5217, 5314 \\ \hline
$m_{\tilde{\nu}_{e}}$,$m_{\tilde{\nu}_{\tau}}$ & {\color{red}559.5}, 2086 & {\color{red}615.1}, 1748 & 4076, 5506 \\
$m_{\tilde{e}_{1}}$,$m_{\tilde{e}_{2}}$ & {\color{red}561.4}, 673.4 & {\color{red}617.1}, 730.5 & 5503, 6325 \\
$m_{\tilde{\tau}_{1}}$,$m_{\tilde{\tau}_{2}}$ & 1600, 2115 & 1550, 1833 & 3709, 4077 \\ \hline
$\sigma_{{\rm SI}}$ & $ 9.6 \times 10^{-14} $ & $ 1.01 \times 10^{-13} $ & $ 8.74 \times 10^{-12} $ \\
$\sigma_{{\rm SD}}$ & $ 1.82 \times 10^{-10} $ & $ 1.95 \times 10^{-10} $ & $ 1.62 \times 10^{-8} $ \\
$\Omega h^{2}$ & 0.103 & 0.121 & 0.122 \\ \hline
$C$ & 0.189 & 0.191 & 0.054 \\ \hline
\end{tabular}}
\caption{A second set of benchmark points which exemplify the QYU scenario compatible with the muon $g-2$ measurements within $2\sigma$. The points are chosen to be consistent with the mass bounds and constraints from rare $B-$meson decays, while they are required not to yield a relic density for LSP neutralino larger than the Planck bounds.}
\label{tab:QYU}
\end{table}



\clearpage

\section{Conclusion}

We have explored some of the phenomenological predictions of a supersymmetric $SU(4)_c \times SU(2)_L \times  SU(2)_R$ model, after taking into consideration the bounds on the dark matter relic abundance and the latest results from the Fermilab muon $g-2$ experiment. If the discrepancy between the SM prediction of muon $g-2$ and the measured value is around $5\sigma$, we identify a variety of coannihilating scenarios involving supersymmetric particles that can resolve it. However, if the muon $g-2$ discrepancy is significantly lower, of order $1\sigma $ or less, a scenario with TeV scale Higgsino dark matter is also viable. Furthermore, this latter scenario is also compatible with third family quasi-Yukawa unification. We display regions of the allowed parameter space that can be probed in the ongoing direct and indirect dark matter detection experiments.

\section*{Acknowledgment}
AT is partially supported by the Bartol Research Institute, University of Delaware. We acknowledge Information Technologies (IT) resources at the University of Delaware, specifically the high performance computing resources for the calculation of results presented in this paper. CSU also acknowledges the resources supporting this work in part were provided by the CEAFMC and Universidad de Huelva High Performance Computer (HPC@UHU) located in the Campus Universitario el Carmen and funded by FEDER/MINECO project UNHU-15CE-2848.

\bibliographystyle{JHEP}
\bibliography{QYU.bib}

\end{document}